\documentclass[aps,prb,reprint,groupedaddress,superscriptaddress,nofootinbib]{revtex4-1}
\usepackage{mathrsfs}
\DeclareMathAlphabet{\mathpzc}{OT1}{pzc}{m}{it}
\usepackage{color}
\usepackage{graphicx}
\usepackage{dcolumn}
\usepackage{bm}
\usepackage{array}
\usepackage{multirow}
\usepackage{epstopdf}

\usepackage{amsmath}
\usepackage{tabulary}
\usepackage{rotating}
\usepackage{graphicx} 
\usepackage{lipsum} 
\usepackage{tabularx}

\usepackage{subfigure}
\usepackage{parskip}

\usepackage{tikz}
\usepackage{amsmath}

\usepackage{float}

\usetikzlibrary{shadings,patterns}
\usetikzlibrary{positioning, shapes.geometric}

\begin{document}
	\preprint{APS/123-QED}
	
	\title{Random Green's function method for large-scale electronic structure calculation}
	\author{Mingfa Tang}
      \affiliation{School of Physical Science and Technology, ShanghaiTech University, Shanghai, 201210, China}
	\author{Chang Liu}	
      \affiliation{Xiaogan Sichuang Information Technology Co., LTD, Xiaogan, Hubei, 432000, China}
    \author{Aixia Zhang}
     \affiliation{School of Physical Science and Technology, ShanghaiTech University, Shanghai, 201210, China}
	\author{Qingyun Zhang}
     \affiliation{School of Physical Science and Technology, ShanghaiTech University, Shanghai, 201210, China}
	\author{Jiayu Zhai}
     \affiliation{Institute of mathematical sciences, ShanghaiTech University, Shanghai, 201210, China}
     \author{Shengjun Yuan}
     \affiliation{Key Laboratory of Artificial Micro- and Nano-structures of Ministry of Education and School of Physics and Technology, Wuhan University, Wuhan 430072, China}
     \affiliation{Wuhan Institute of Quantum Technology, Wuhan, Hubei 430206, China}
	\author{Youqi Ke}
	\email{keyq@shanghaitech.edu.cn}
	\affiliation{School of Physical Science and Technology, ShanghaiTech University, Shanghai, 201210, China}
	\date{\today}

	\begin{abstract}
 We report a Krylov projected random Green's function (rGF) method for linear-scaling electronic structure calculation. In this method, the rGF is defined on a set of random states and is calculated by projecting onto Krylov subspace to obtain Fermi-Dirac operator efficiently. To demonstrate the applicability, we implement rGF method with density-functional Tight-Binding method. We show the Krylov subspace can maintain at small size for gaped materials at low temperature T. We find, with a simple deflation technique, the rGF self-consistent calculation of $H_2O$ clusters at $T=0$ K can reach an error of $\sim$1meV per $H_2O$ molecule in total energy, compared to deterministic calculations. The rGF method provides an effective stochastic method for large-scale electronic structure simulation.
%
	\end{abstract}
	\pacs{Valid PACS}
	\maketitle

An accurate and linear-scaling electronic structure (ES) method is essential for large-scale applications, and has being highly pursued in the materials simulation community.\cite{1999ReviewModernPhysics,Bowler2012} The standard implementation of Kohn-Sham density functional theory (KSDFT)\cite{HohenbergPKohnWDFT1964, KohnWandShamLJDFT1965} is difficult to deal with systems more than hundreds or thousands of atoms, due to the cubic-scaling computational complexity of diagonalization. To circumvent the diagonalization, many classic linear-scaling methods have been developed by employing various approximations based on the Kohn's $nearsightedness$ principle,\cite{VanderbiltPhysRevB1993} including divide-and-conquer\cite{YangPhysRevLett1991} and other approximate density matrix (DM) techniques.\cite{VanderbiltPhysRevB1993, VanderbiltPhysRevB1994,NiklassonPRB2002, SGoedeckerandMTeterPRB1995} Nonetheless, as cost for achieving linear scaling, these O(N) ES methods suffer from the large prefactor and the limited accuracy, and thus present limitations on the dimensionality and the character of systems.\cite{1999ReviewModernPhysics,Bowler2012} 
Recently, the random-state method\cite{Alben1975} has presented a new avenue for developing linear-scaling (or even sub-linear) simulations of materials' ES, including the reports of stochastic DFT\cite{BaerPhysRevLett2013}, and time-dependent random-state propagation based DFT\cite{YuanCPL2023}. The random-state first-principles simulation of over $10^4$ atoms have been reported.\cite{YuanCPL2023} It has been proven that the random-state method bears only the unbiased distributed statistical error of the order $1/\sqrt{N_s}$ ($N_s$ is the number of random states), and the error in the averaged global property, e.g. energy per electron, scales as $1/\sqrt{N_eN_s}$ ($N_e$ is the electron number) due to the self-average.\cite{AnthonyHamsPRE2000} Various techniques, e.g. the deflation\cite{OrginosKostasSIAMJournalonScientificComputing2017} and probing\cite{YousefSaadNumericalLinearAlgebrawithApplications2012, AndreasStathopoulosandJesseLaeuchliandKonstantinoOrginoSIAMJSciComput2013, CristianDJournalofChemicalPhysics2018, AndrewDBaczewskiElectronicStructure2019, AndreasSIAMJournalonScientificComputing2020} approaches, exist for effectively reducing the statistical errors, providing the important basis for applying random-state method for ES applications. It has been shown that the statistical error in random-state based DFT can be effectively reduced by about an order of magnitude by combining with the techniques of embedded fragment,\cite{BaerJCP2014, BaerJCP2017,BaerJCP2019} energy windows,\cite{BaerJCP2021}a universal mixed stochastic-deterministic approach\cite{CollinsPhysRevLett2020}, to achieve important accuracy in charge density, energy, and forces. The present implementation of random-state based DFT utilizes the Chebyshev polynomial expansion to represent the Fermi-Dirac operator,\cite{BaerPhysRevLett2013,YuanCPL2023, BaerWIREsComputationalMolecularScience2019} is thus suitable for high temperature. However, for low-temperature narrow-gap systems, a large number of Chebyshev polynomials is required for high accuracy.\cite{BaerPhysRevLett2013,YuanCPL2023} 
It is noted that a recent report of the rational approxiamtion for implementing random-state methods presents the computational cost similar to the polynomial approximation for gold metal clusters at finite temperature.\cite{AndrewDBaczewskiElectronicStructure2019}  
In this work, we present a linear-scaling random Green's function (rGF) method for large-scale ES simulation by projecting onto a small Krylov subspace to calculate the Fermi-Dirac operator efficiently. 

 To start, a general rGF $\tilde{G}$ can be introduced as,
\begin{equation}
\tilde{G}=G\cdot\tilde{I},
\end{equation}
where $G$ is the exact GF which models the system's spatial and temporal response possessing wide applications in quantum physics. For example, the Keldysh's GF $G^{\mathcal{K}}$ and the retarded GF $G^{\mathcal{R}}$ are the central quantities for simulating the respective nonequilibrium and equilbrium systems.\cite{Mahanmanybodytheory,KadanoffQuantumstatisticalmechanics} Here, $\tilde{I}$ is an approximate random resolution of identity matrix,\cite{BaerPhysRevLett2013, HutchinsonCommunications_in_Statistics1990} given by 
\begin{equation}
\tilde{I}=\frac{1}{N_s}\sum_{i=1}^{N_s}| X_i \rangle \langle X_i |,
\end{equation}
where  $| X_i \rangle$ is the random state with the random elements $1$ or $-1$ of equal probability (other choice of $\tilde{I}$ is possible.), $N_s$ is the number of random states. It is clear that $\tilde{I}=I+\delta$ with $\delta_{ii}=0$. The random $\delta_{ij} (i\neq j)$ satisfies the Gaussian distribution (due to the central limit theorem) with the average $\langle \delta_{ij} \rangle=0$, and  a standard deviation   $\langle \delta^2_{ij} \rangle = 1/N_s$. The randomization of GF has been applied to accerlerate GW calculations.\cite{GWPRLstochastic} In this work, we present the application of rGF for simulating large-scale ES with the efficient Krylov projection method.

For equilibrium noninteracting ES, we consider a physical system with the overlap $O$ (for generality) and Hamiltonian $H$ defined with $N$ localized bases. Then, a randomized matrix function, $\tilde{F}(O^{-1}H)$, e.g. Fermi-Dirac function $F(X)=1/(exp(\beta(X-\mu))+1)$ ($\beta=1/k_BT$, $\mu$ is chemical potential) can be connected to the rGF by,
\begin{equation}\label{DM1}
\tilde{F}=-\frac{1}{\pi}Im\{O\int^{+\infty}_{-\infty} F(\epsilon) \tilde{G}^{\mathcal{R}}(z)d\epsilon\}=F\tilde{I}
\end{equation}
where the rGF $\tilde{G}^{\mathcal{R}}(z)$ is given by the retarded GF $G^{\mathcal{R}}(z)=[zO-H]^{-1}$ ($z=\epsilon+i\eta$,$\eta$ is infinitesimal). $\tilde{F}$ bears only the statistical error, namely $F\cdot\delta$, which is controllable with the finite $N_s$. For a large scale system, $N_s$ presenting a good accuracy can be much smaller than the dimension of $H$. By introducing the polynomial expansion to $F({\epsilon})$ in Eq.\ref{DM1}, one can obtain the polynomial approximation of $\tilde{F}(O^{-1}H)$, e.g. the Chebyshev expansion for the DM,\cite{BaerPhysRevLett2013,YuanCPL2023} avoiding the complexity for calculating rGF and the integration. However, Polynomial expansion of Fermi-Dirac operator requires high orders for narrow-gap materials at low T. In this work, we develop a distinct approach by calculating the rGF and its associated integration with linear scaling computation.

To proceed, we rewrite the retarded rGF as 
\begin{equation}\label{rGF1}
\tilde{G}^{\mathcal{R}}(z) = \frac{1}{N_s}\sum_{i=1}^{N_s} | R_i(z) \rangle \langle X_i |
\end{equation}
where the vector $|R_i(z)\rangle$ satisfies the linear equation 
\begin{equation}\label{LinearEQGF}
 [zO-H]|R_i(z)\rangle=|X_i\rangle.
\end{equation}
which features a $shifted$ complex symmetric linear system for different z which can be solved by Krylov subspace methods.\cite{YamamotoPhysRevB2011, KrylovSubspace, SLZhangKrylovSubspacediagonal2012}. For applications for many $z$ points and the rGF integration in Eq.\ref{DM1}, we can adopt a common Krylov subspace,\cite{KrylovSubspace,SLZhangKrylovSubspacediagonal2012} which is constructed with the Arnoldi process,\cite{KrylovSubspace} denoted by $K^{(i),v} = K_{v}(H',|\mathcal{X}_i\rangle)$,
where $v$ is the size of subspace, $H'=O^{-1}H$ and $|\mathcal{X}_i\rangle=O^{-1}|X_i\rangle$. Here, $K^{(i),v}$ contains $\{|\Phi^{(i),v}_m\rangle,m=1,...,v\}$ in which the involved $O^{-1}$-vector product is solved efficiently with the iterative conjugate gradient (CG) linear solver. Then, we can obtain a projected $v \times v$ subspace overlap $O^{(i),v}_{K,mn}=\langle \Phi^{(i),v}_m|O|\Phi^{(i),v}_n\rangle$ and  Hamiltonian $H^{(i),v}_{K,mn}=\langle \Phi^{(i),v}_m|H|\Phi^{(i),v}_n\rangle$. Then, in $K^{(i),v}$,  the subspace GF is obtained $\tilde{G}^{(i),v}_K(z)=(zO^{(i),v}_K-H^{(i),v}_K)^{-1}$, then
\begin{equation}\label{rGF2}
|R_i(z)\rangle = \sum^{v}_{m,n}G^{(i),v}_{K,mn}(z) |\Phi^{(i),v}_m\rangle  \langle\Phi^{(i),v}_n|X_i\rangle,
\end{equation}
to give rGF in Eq.\ref{rGF1}. As a result, by combining Eq.\ref{rGF1} and Eq.\ref{rGF2}, Eq.\ref{DM1} is projected onto a Krylov subspace integration, namely 
\begin{equation}\label{KrylovGFInte}
\tilde{F}^{(i),v}_{K}= -\frac{1}{\pi}Im[\int dz F(z) \tilde{G}_K^{(i),v}(z)],
\end{equation}
which can be obtained analytically with subspace diagonalization or numerically with accurate complex-contour technique (based on the analyticity of retarded GF).\cite{Turrekeybook} Finally, we can obtain
\begin{equation}\label{rGF3}
O^{-1}\tilde{F}=\frac{1}{N_s}\sum^{N_s}_{i}\sum^{v}_{m,n}\tilde{F}_{K,mn}^{(i),v}|\Phi^{(i),v}_m\rangle  \langle\Phi^{(i),v}_n|X_i\rangle\langle X_i |.
\end{equation}
 It should be noted that Eq.\ref{rGF3} essentially provides a Krylov expansion with $v$ polynomials of $H'$ (distinct from the Chebyshev expansion), featuring a linear scaling computational cost. As an application of rGF, the generalized non-interacting DM, using Fermi-Dirac function in Eq.\ref{DM1}, can be obtained as  $\tilde{\mathcal{D}}=O^{-1}\tilde{F}=\mathcal{D}\tilde{I}$ ($\mathcal{D}=O^{-1}F$ is the exact DM). The Krylov subspace method provides a controllable approach to calculate $\tilde{D}$ with high accuracy. The convergence of the Krylov subspace for $\tilde{D}$ can be related to the number of distinct clusters of eigenvalues of the material system and their width.\cite{meyerkrylov} Moreover, for the linear-algebraic feature of the rGF method, a large volume of techniques in linear algebra, e.g. various preconditioning techniques, can be combined to further improve the efficiency.
For the converged $v$, the DM $\mathcal{D}$ only bears a statistical error, namely $\mathcal{D}\cdot\delta$ scaling as the $1/\sqrt{N_s}$. For an effective error reduction, the general deflation and probing approach can be applied.\cite{OrginosKostasSIAMJournalonScientificComputing2017, YousefSaadNumericalLinearAlgebrawithApplications2012, AndreasStathopoulosandJesseLaeuchliandKonstantinoOrginoSIAMJSciComput2013, CristianDJournalofChemicalPhysics2018, AndrewDBaczewskiElectronicStructure2019, AndreasSIAMJournalonScientificComputing2020}

To demonstrate the applicability of the rGF method, we implement it with the density functional Tight-Binding approach (DFTB)\cite{KaschnerPRB1995, FrauenheimIJOQC1996, FrauenheimPRB1998, MakinenComputationalMaterialsScience2009, SeifertPhysicalandEngineeringSciences2014} to calculate the ES and total energy and compare with the deterministic method. We use the non-orthogonal basis which usually possesses the higher accuracy and transferability than orthogonal basis. In each DFTB self-consistent iteration, the Mulliken charge $q_R$ for site $R$, required for updating the Hamiltonian $H$, is calculated as, 
\begin{equation}\label{eq10}
q_{R}= Tr(\{\tilde{\mathcal{D}}\cdot O\}_{RR}).
\end{equation}
In DFTB, the total orbital energy is obtained as
\begin{equation}
 E_{orbit}=Tr(\tilde{ \mathcal{D}}\cdot H), 
\end{equation}
and then the DFTB total energy is given by,
\begin{equation}
 E_{tot}=E_{orbit}+E_{rep}-E_{coul}, 
\end{equation}
where $E_{rep}$ is pairwise energy in which the parameters are fitted to the high-level theoretical calculation(e.g. KSDFT), and $E_{coul}$ is coulomb energy that corrects the double counting in $E_{orbit}$.\cite{KaschnerPRB1995, FrauenheimIJOQC1996, FrauenheimPRB1998, MakinenComputationalMaterialsScience2009, SeifertPhysicalandEngineeringSciences2014} In the rGF-DFTB calculations, to ensure the convergence of the self-consistent procedure, the same set of random states are used in all the iterations, avoiding the stochastic noise. For applications, all the codes are compiled with the intel Math Kernel Library, and we use the CG solver package PETSc\cite{PETScwebside, PETScUsersManual, PETScBook} for $O^{-1}$-vector product and matrix-vector multiplication in constructing the Krylov subspace. The present rGF-DFTB solver is implemented based on the open-source package LATTE. All the calculations are performed on the Intel Xeon Gold 6226 CPU (2.70GHz). We adopt exact diagonalization for the Krylov subspace integration in Eq.\ref{KrylovGFInte} for the different T.\cite{supplement}

\begin{figure}[!htbp]
	 \centering
	\includegraphics[width=1.0\columnwidth]{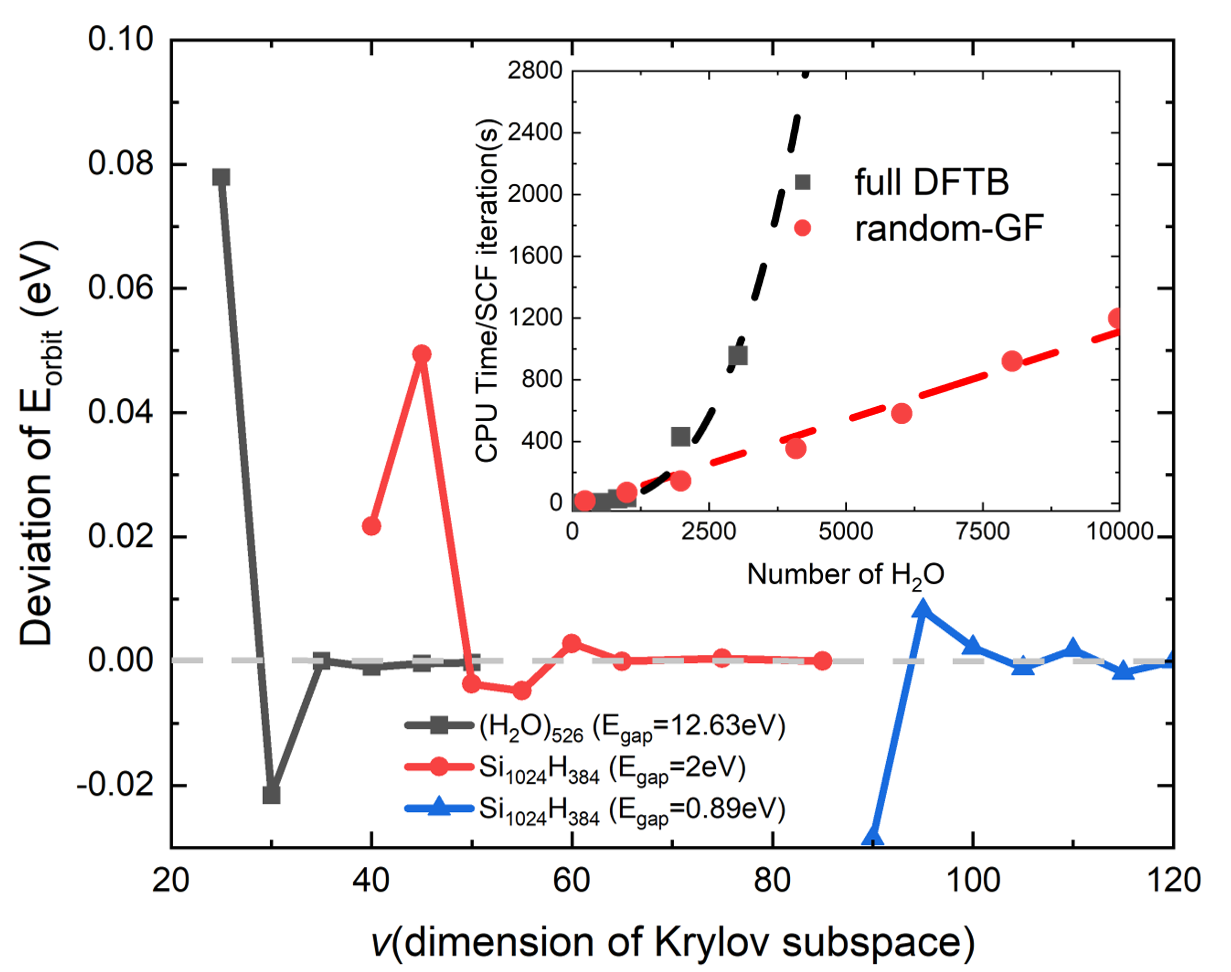}
	\caption{Deviation of $E_{orbit}$ versus $v$ for $(H_2O)_{526}$, $Si_{1024}H_{384}$ and $\overline{Si}_{1024}H_{384}$ clusters at $T=0$. 
 Inset: total CPU time per iteration for the full-DFTB and rGF ($N_s=1000$) calculations performed on a single CPU core.}
	\label{figure1}
\end{figure}

In Fig.\ref{figure1}, we investigate the convergence of Krylov subspace, namely deviation of $E_{orbit}$ versus $v$ ($N_s=1000$), for $(H_2O)_{526}$, $Si_{1024}H_{384}$,and $\overline{Si}_{1024}H_{384}$ clusters at $k_BT=0 eV$ (based on the self-consistent charges of full DFTB in LATTE).\cite{LATTEwebsite}  We use the minimal 6 and 4 atomic bases for the respective $H_2O$ and Si with the associated DFTB parameters as in Refs.\onlinecite{CawkweJChemTheoryComput2017, FrauenheimSolidStateCommunications1999, BiegerZPhysChemie1986}. $H_2O$ clusters are generated with the solvation box module in VMD,\cite{VMD} and $Si$- and $\overline{Si}$-clusters are cut from the Diamond bulk with the respective lattice constants $3.86 \overset{\circ}{A}$ and  $4.06 \overset{\circ}{A}$. The HOMO-LUMO energy gaps are 12.63 eV, 2.0 eV and 0.89eV for the respective $H_2O$,$Si$ and $\overline{Si}$ clusters. It is found that $E_{orbit}$ can be converged with an error $\le 5meV$ with $v=35,65$ and $100$ for the respective $H_2O$,$Si$ and $\overline{Si}$ clusters. However, as shown in supplement,\cite{supplement} to safely converge the energy within few meV at $k_BT=0.03eV$, Chebyshev expansion requires 2000 polynomials for $H_2O$ and 1400 polynomials for both $Si_{1024}H_{384}$ and $\overline{Si}_{1024}H_{384}$. By increasing temperature to $k_BT=0.1$, the required number of Chebyshev polynomials can be significantly reduced to 600, 300 and 400 for the respective $H_2O$, $Si$ and $\overline{Si}$ clusters, which are still much higher than the Krylov polynomials $v$ at $T=0$.  In addition,it is also noted that the increase of temperature can also reduce the number of $v$ for the three clusters (see the supplement). The small size of Krylov subspace $v$ provides the important efficiency for large-scale ES simulations with the rGF method.

\begin{figure}[!htbp]
	 \centering
	\includegraphics[width=1.0\columnwidth]{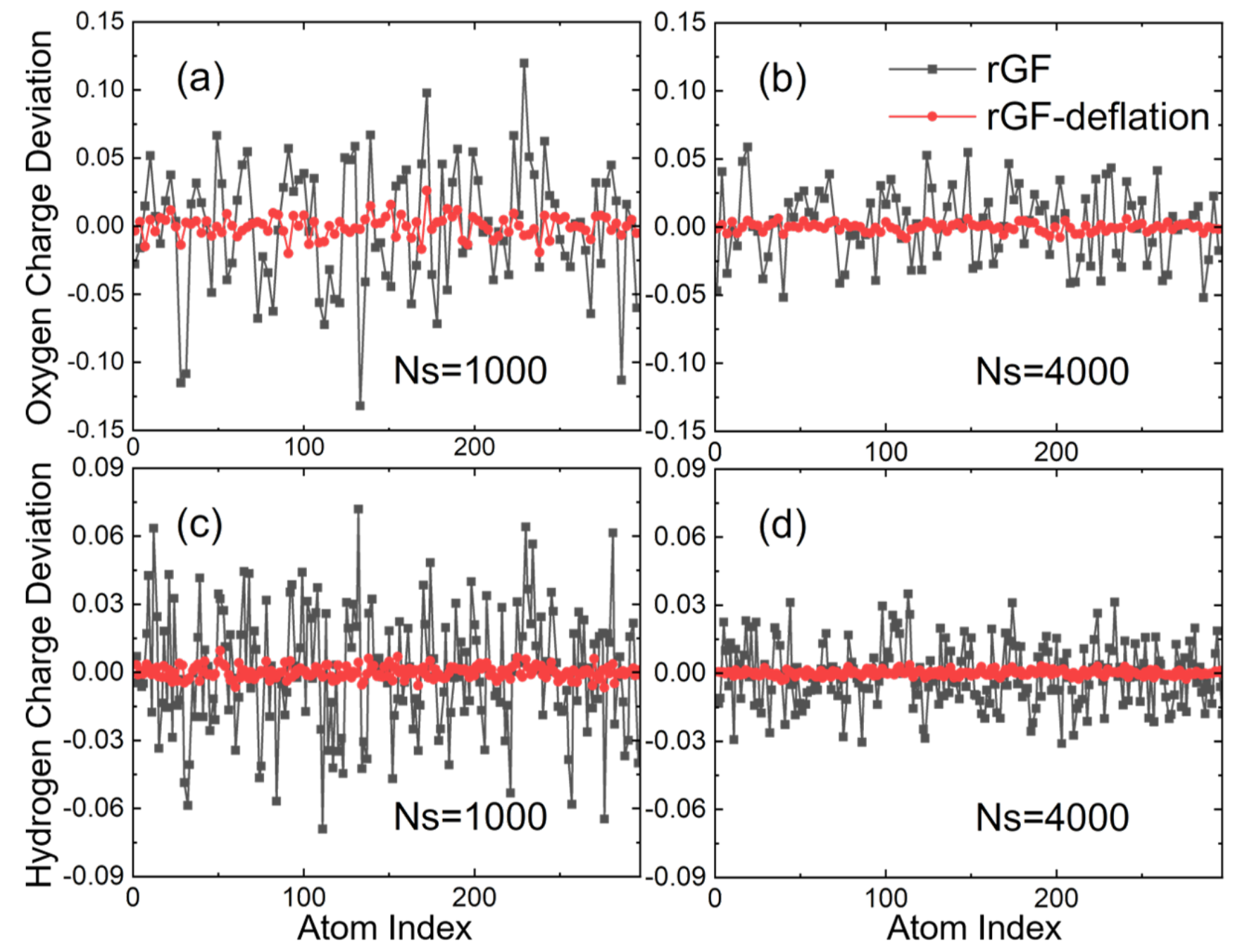}
	\caption{Stochastic charge deviation from full DFTB calculation for $(H_2O)_{99}$ cluster: (a) and (c)are for $N_s=1000$, (b) and (d) are for $N_s=4000$, the results with (circles) and without (squares) deflation correction are compared for H (c,d) and O (a,b) atoms.}
	\label{figure2}
\end{figure}

\begin{table*}[!htbp] 
  \centering 
  \caption{Total energy per $H_2O$ for the deterministic full DFTB, rGF and rGF-deflation calculations (with $N_s=$1000 and 4000) for different $H_2O$ clusters. } 
  \label{tab:example} 
  \begin{tabularx}{\textwidth}{XXXXXX}
\hline\hline
		& $E/H_2O$   \\
	system	& full-DFTB & rGF & rGF  & rGF-deflation & rGF-deflation\\
	$N_s$& & 1000 & 4000  & 1000 & 4000\\
\hline
 	$(H2O)_{99}$ 	&-110.4922 	&-110.3071	    &-110.3428    & -110.4904 &-110.4926     \\	
	$(H2O)_{233}$	&-110.5078	    &-110.3473	    & -110.4652  &-110.4976  & -110.5065 \\		
	$(H2O)_{526}$   &-110.5069   & -110.4630   &-110.4927   & -110.5070 & -110.5042 \\	
	$(H2O)_{995}$ 	&-110.5103    & -110.4202   & -110.4967 & -110.5130 &   -110.5114\\	
	$(H2O)_{1981}$	 &-110.5153   &-110.4860    &-110.5162   &-110.5173  & -110.5165    \\	
	$(H2O)_{4087}$	& \textemdash     &\textemdash 	& \textemdash  	&-110.5206   &   -110.5203   \\	
	$(H2O)_{9984}$ & \textemdash 	& \textemdash 	& 	\textemdash 	& -110.5229   & -110.5231   \\	 
  \hline\hline
  \end{tabularx}
\end{table*}

In the next, we investigate the accuracy of self-consistent rGF ES calculations for $H_2O$ clusters with different sizes,in comparison with the deterministic full DFTB calculation. To effectively reduce the statistical error in rGF method, we adopt the deflation technique (denoted by rGF-deflation), namely $\tilde{\mathcal{D}}= \mathcal{D}_0 + (\mathcal{D}-\mathcal{D}_0)\tilde{I}$,
where $\mathcal{D}_0$ denotes a good sparse approximation to $\mathcal{D}$. We use the $ D_{H_2O}$ of each single $H_2O$ in the cluster to construct a sparse $\mathcal{D}_0$, and keep it fixed in the self-consistency. All self-consistent calculations are converged to the tolerance $10^{-5}$ in $q_R$. Fig.\ref{figure2} presents the mulliken charge deviation $\Delta q_R$ from the full DFTB results for both rGF and rGF-deflation calculations of $(H_2O)_{99}$ cluster, and the results for $N_s=1000$ and $4000$ are compared. As shown, for both O and H atoms, rGF-deflation results present almost one order of magnitude  smaller than the rGF calculations, presenting the important correction to rGF ES calculations. In particular, for $N_s=1000$,  rGF calculation presents the values of averaged charge deviation, namely $\langle |\Delta q| \rangle=\sum_{R=1}^{N_{Atom}}|\Delta q_{R}|/N_{Atom}$,  $0.02$ and $0.038$ for the respective H and O atoms, compared to the corresponding values $0.002$ and $0.006$ of rGF-deflation results. By increasing $N_s$ to 4000, the charge deviation $\langle |\Delta q| \rangle$ is reduced to $0.01$ and $0.02$  with rGF, and $0.001$ and $0.0025$ with rGF-deflation for the respective H and O atoms. By comparing the $N_s=1000$ and $N_s=4000$ results of both rGF and rGF-deflation calculations, we can see the statistical error present a $1/\sqrt{N_s}$ dependence. In addition, the statistical error in both rGF and rGF-deflation results show unbiased distribution throughout the water cluster. Nevertheless, in percentage, $\langle |\Delta q| \rangle / q_R$  is rather small, for example, for O atom, only  0.6$\%$ and 0.09$\%$ with the respective rGF and rGF-deflation of $N_s=1000$ (even smaller with $N_s=4000$).

 In table.\ref{tab:example},  we present the total energy per $H_2O$ for different clusters self-consistently calculated by rGF, rGF-deflation with $N_s=1000$ and $4000$, together with the benchmark full DFTB results for comparison. Compared to full DFTB, rGF ($N_s=1000$) results present large error in the total energy, for example, 185meV and 160meV/$H_2O$ in the respective $(H_2O)_{99}$ and $(H_2O)_{233}$. 
One can also easily find that the total energy error for rGF ($N_s=1000$)  presents the important self-average by the cluster size, namely the error decreases with increasing the cluster size (for example, the error is reduced to 29 $meV/H_2O$ in $(H_2O)_{1981}$), consistent with the Refs.\onlinecite{BaerPhysRevLett2013,YuanCPL2023}. The large total energy error in rGF ($N_s=1000$) results is consistent with the large charge deviation as shown in Fig.\ref{figure2}. However, compared to  rGF ($N_s=1000$) calculations, the rGF ($N_s=4000$) results present significant reduction in the total energy error, presenting better agreement with the full DFTB.  For example, the energy errors in rGF($N_s=4000$) are $42.6meV$ and $14.2meV$/$H_2O$ in the respective $(H_2O)_{233}$ and $(H_2O)_{526}$, much smaller than the values $160meV$ and $44meV$/$H_2O$ meV of rGF ($N_s=1000$). When increasing cluster size to $(H_2O)_{1981}$, the energy error in rGF($N_s=4000$) is decreased to 5.9meV/$H_2O$, presenting the important self-average effect. When it comes to the rGF-deflation ($N_s$=1000) calculations, we find the error in total energy is  few meV,for example, only 1.8 meV/$H_2O$ for the cluster $(H_2O)_{99}$, which is dramatic improved over the rGF calculations. By increasing $N_s$ to $4000$, the total energy deviation from full DFTB calcuation falls to $\sim$1meV or even smaller, demonstrating the important effectiveness of deflation. In addition, it is found that, as increasing the cluster size to $(H_2O)_{4086}$ and $(H_2O)_{9984}$,  rGF-deflation calculations with $N_s$=1000 and 4000 can present a deviation less than 1 meV. With better choices of $\mathcal{D}_0$\cite{VanderbiltPhysRevB1993, YangPhysRevLett1991, NiklassonPRB2002, SGoedeckerandMTeterPRB1995} or applying the probing technique,\cite{YousefSaadNumericalLinearAlgebrawithApplications2012, AndreasStathopoulosandJesseLaeuchliandKonstantinoOrginoSIAMJSciComput2013, CristianDJournalofChemicalPhysics2018, AndrewDBaczewskiElectronicStructure2019, AndreasSIAMJournalonScientificComputing2020} the accuracy of rGF is expected to be further improved. 

Finally, the inset of Fig.\ref{figure1} shows the total CPU time per iteration versus system size for the rGF-deflation ($N_s=1000$) and full DFTB calcualtions. We can find that the CPU time of rGF method scales linearly with system size, while full DFTB presents almost cubic scaling. The CPU time of rGF-deflation intersects with that of full DFTB at about $(H_2O)_{1500}$ and $(W)_{1500}$. For calculating the $(H2O)_{9984}$ cluster, the rGF takes about 1200 seconds on a single CPU core. For the rGF calculations, significant acceleration is expected by introducing effective preconditioners\cite{NiklassonPRB2004, NiklassonAndersMauritzJCTC2016} and developing the multiple Krylov space as reported in Ref.\onlinecite{SLZhangKrylovSubspacediagonal2012} to avoid the CG loop $O^{-1}H$-vector product in constructing Krylov subspace for non-orthogonal basis.

In summary, we have developed a linear scaling rGF method for simulating large-scale materials. The rGF is defined on a set of random states and is projected onto a small Krylov subspace to calculate the DM. By combining with the DFTB, we have demonstrated the rGF self-consistent electronic structure calculation can reach an error of $\sim$1meV per $H_2O$ in total energy, in comparison with the deterministic calculation. The random Green's function is generally applicable to many other electronic structure methods, including density functional theory, Tight-Binding model and nonequilibrium Green's function method for nanoelectronics, quantum many-body method.

Y.Ke acknowledges financial support from NSFC (grant No.12227901), and S.Yuan.thanks the financial support from NSFC(grant No.11974263,12174291). The authors thank the HPC platform of ShanghaiTech University for providing the computational facility.

	\bibliographystyle{apsrev4-2}
	\bibliography{ref}

\end{document}